\documentclass[12pt]{article}
\usepackage{amsmath,amssymb,bm,graphicx}
\begin{document}

\vskip 0.5 truecm
\begin{center}
{{\Large{\bf  Heisenberg Uncertainty Relation Revisited}}\\
\vspace{0.5cm}
---{\large{\it Universality of Robertson's relation ---}}}
\end{center}

\vskip .5 truecm
\centerline{\bf  Kazuo Fujikawa}
\vskip .4 truecm
\centerline {\it RIKEN Nishina Center, Wako 351-0198, Japan}
\vskip 0.5 truecm

\makeatletter
\@addtoreset{equation}{section}
\def\theequation{\thesection.\arabic{equation}}
\makeatother

\begin{abstract}
It is shown that all the known uncertainty relations are the secondary consequences of Robertson's relation. The basic idea is to use the Heisenberg 
picture so that the time development of quantum mechanical operators incorporate the effects of the measurement interaction. A suitable use of triangle inequalities then gives rise to various forms of uncertainty relations. The assumptions of unbiased measurement and unbiased disturbance are important to simplify the resulting uncertainty relations and to give the familiar uncertainty relations such as a naive Heisenberg error-disturbance relation. These simplified uncertainty relations are however valid only conditionally. Quite independently of uncertainty relations, it is shown that the notion of precise measurement is incompatible with the assumptions of unbiased measurement and unbiased disturbance. We can thus naturally understand the failure of the naive Heisenberg's error-disturbance relation, 
as was demonstrated by the recent spin-measurement by J. Erhart, et al..
\end{abstract}

\makeatletter
\@addtoreset{equation}{section}
\makeatother


\section{Introduction}
The Born probability interpretation of quantum mechanics is formulated by~\cite{neumann} 
\begin{eqnarray}
\langle P_{k}\rangle={\rm Tr}\rho P_{k}\geq 0
\end{eqnarray}
which gives the probability of observing the projection operator $P_{k}$
with $\sum_{k}P_{k}=1$. The observed value of any hermitian operator $A$ is thus given by
\begin{eqnarray}
\langle A\rangle={\rm Tr}\rho A\nonumber
\end{eqnarray}
or $\langle A\rangle=\langle\psi|A|\psi\rangle$ for a pure state, and the non-vanishing dispersion (standard deviation)
\begin{eqnarray}
\sigma(A)=\sqrt{\langle (A-\langle A\rangle)^{2}\rangle}\ \neq 0
\end{eqnarray}
in general. Namely, the observed values are not dispersion-free in general in the above sense. This property combined with commutator algebra are the basis of uncertainty relations.

It is well-known that Heisenberg~\cite{heisenberg} proposed his uncertainty relation on the
basis of a gedanken experiment of the measurement of the position and momentum of the electron.  
To define the coordinate within $\Delta x$(= wave length of X-ray, for example), we have the momentum change of the electron
$\Delta p\simeq \hbar/\Delta x$.
We thus define 
\begin{eqnarray}
&&{\rm "Error"}:  \epsilon (x)=\Delta x,\nonumber\\
&&{\rm "Disturbance"}: \eta(p)=\Delta p
\end{eqnarray}
and we have the error-disturbance relation
\begin{eqnarray}
\epsilon (x)\eta(p)\sim \hbar.
\end{eqnarray}

\noindent{\bf Kennard and  Robertson relation:}\\

Immediately after Heisenberg, 
Kennard~\cite{kennard} and  Robertson~\cite{robertson} formulated the uncertainty relation which appears in any textbook on quantum mechanics.
Starting  with hermitian operators which satisfy $[A, B]=iC$, we obtain
the positive definite quantity
\begin{eqnarray}
&&\langle [(A-\langle A\rangle)+is (B-\langle B\rangle)]^{\dagger}
[(A-\langle A\rangle)+is (B-\langle B\rangle)]\rangle\nonumber\\
&&= \langle (A-\langle A\rangle)^{2}\rangle+ s^{2}\langle (B-\langle B\rangle)^{2}\rangle - s \langle C\rangle\geq 0.\nonumber
\end{eqnarray}
The discriminant of this quadratic form of the real variable $s$ gives 
\begin{eqnarray}
\sigma(A)\sigma(B)\geq \frac{1}{2}|\langle [A, B]\rangle|.
\end{eqnarray}

The "conventional" physical interpretation of Robertson's relation (1.5) is that we first measure the variable $A$ repeatedly for an ensemble of similarly prepared quantum mechanical samples  
without paying attention to the variable $B$ and we evaluate 
$\sigma(A)$. Similarly, we measure $\sigma(B)$ for the similarly prepared samples without paying attention to the variable $A$. We then obtain the above uncertainty relation (1.5). We shall present a different interpretation of the general form of Robertson's relation later.
\\

\noindent{\bf Arthurs-Kelly relation:}\\

Another important development in the history of uncertainty relations is
the analysis of Arthurs and Kelly~\cite{arthurs, arthurs2}.
They introduce the measuring apparatus $M$ for $A$, and $N$ for $B$, respectively, with
\begin{eqnarray}
[M,N]=0.
\end{eqnarray}
The notion of unbiased measurement is important in their analysis, which is defined by 
\begin{eqnarray}
\langle M^{out}\rangle=\langle A\rangle
\end{eqnarray}
for {\em any} state of the system $\psi$ in the total Hilbert space of the system and apparatus $|\psi\rangle\otimes|\xi\rangle$ in the Heisenberg picture.
Here variables $M$ and $N$ (and also $A$ and $B$) stand for the variables before the measurement, and the variable $M^{out}=U^{\dagger}MU$ stands for the apparatus $M$ after measurement. Note that measurement is a unitary time development generated by $U$ in this formulation.

As for the unbiased measurement, one can show a much stronger consequence of the assumption such as 
\begin{eqnarray}
\langle (M^{out}-A)B\rangle=\langle B(M^{out}-A)\rangle=0.
\end{eqnarray}
This stronger result follows from the identity~\cite{appleby}
\begin{eqnarray}
\langle\psi\otimes\xi|{\cal A}|\psi^{\prime}\otimes\xi\rangle
&=&
\frac{1}{4}\{
\langle(\psi+\psi^{\prime})\otimes\xi|{\cal A}|(\psi+\psi^{\prime})\otimes\xi\rangle\nonumber\\
&&-\langle(\psi-\psi^{\prime})\otimes\xi|{\cal A}|(\psi-\psi^{\prime})\otimes\xi\rangle\nonumber\\
&&-i\langle(\psi+i\psi^{\prime})\otimes\xi|{\cal A}|(\psi+i\psi^{\prime})\otimes\xi\rangle\nonumber\\
&&+i\langle(\psi-i\psi^{\prime})\otimes\xi|{\cal A}|(\psi-i\psi^{\prime})\otimes\xi\rangle\}
\end{eqnarray}
with ${\cal A}=M^{out}-A$ and $\psi^{\prime}=B\psi$, for example. Note that we have only "diagonal" elements in (1.9).

By noting $M^{out}=M^{out}-A+A$ and using the assumption of unbiased measurements of $A$ and $B$, one obtains 
\begin{eqnarray}
&&\sigma(M^{out})^{2}= \sigma(M^{out}-A)^{2}+\sigma(A)^{2}, \nonumber\\
&&\sigma(N^{out})^{2}= \sigma(N^{out}-B)^{2}+\sigma(B)^{2}.
\end{eqnarray}
If one notes the Heisenberg error-error relation (the derivation of this relation will be given later)
\begin{eqnarray}
\epsilon(M^{out}-A)\epsilon(N^{out}-B)\geq \frac{1}{2}|\langle [A, B]\rangle|,
\end{eqnarray}
where the error is defined by  
\begin{eqnarray}
\epsilon(M^{out}-A)\equiv\langle (M^{out}-A)^{2}\rangle^{1/2}\geq \sigma(M^{out}-A),
\end{eqnarray}
and the standard Robertson's relation (1.5), we  have 
\begin{eqnarray}
\sigma(M^{out})^{2}\sigma(N^{out})^{2}
&\geq& \frac{1}{4}|\langle [A, B]\rangle|^{2}\{[\sigma(N^{out}-B)^{2}]^{-1}+[\sigma(B)^{2}]^{-1}\}\nonumber\\
&&\times\{\sigma(N^{out}-B)^{2}+\sigma(B)^{2}\}\geq
|\langle [A, B]\rangle|^{2}.
\end{eqnarray}
We thus obtain the {\em Arthurs-Kelly relation}
\begin{eqnarray}
\sigma(M^{out})\sigma(N^{out})\geq |\langle [A, B]\rangle|\nonumber
\end{eqnarray}
for unbiased joint measurements. A salient feature of this relation is that 
the left-hand side contains only the standard deviations of measuring apparatus, and the lower bound is twice of the common relation~\cite{she, yuen,yamamoto, braustein, stenholm}.\\

\noindent{\bf Naive Heisenberg error-disturbance relation:}\\

Traditionally, it has been common to take the relation~\cite{braginsky, appleby,ozawa1}
\begin{eqnarray}
\epsilon(A)\eta(B)\geq \frac{1}{2}|\langle [A,B]\rangle|
\end{eqnarray}
as the naive Heisenberg error-disturbance relation; we use the adjective  
"naive" since no reliable derivation of this relation is known.
Here we defined the {\em error} of $A$ by
\begin{eqnarray}
\epsilon(A)&\equiv&\langle (M^{out}-A)^{2}\rangle^{1/2}\geq \sigma(M^{out}-A)
\end{eqnarray}
and the {\em disturbance} of $B$ by 
\begin{eqnarray}
\eta(B)&\equiv&\langle (B^{out}-B)^{2}\rangle^{1/2}\geq\sigma(B^{out}-B)
\end{eqnarray}
with $B^{out}=U^{\dagger}BU$.\\
  
\noindent {\bf Recent experimental test:}\\

An elegant experiment of spin measurement by J. Erhart et al.~\cite{hasegawa}, invalidated  
the naive Heisenberg-type error-disturbance relation (1.14), which initiated the recent activities on uncertainty relations. See also the experiments in~\cite{exp2,exp3,exp4,exp5,exp6}.
In their experiment,  the non-commuting variables $A=\sigma_{x}$, 
$B=\sigma_{y}$ and the state $|\psi\rangle=|+z\rangle$ was used.
Their experiment also confirmed the validity of a modified form of error-disturbance relation 
\begin{eqnarray}
\epsilon(A)\eta(B)+\sigma(A)\eta(B)+\epsilon(A)\sigma(B)\geq \frac{1}{2}|\langle [A,B]\rangle|
\end{eqnarray}
which was proposed some time ago by Ozawa~\cite{ozawa1}. But the lower bound of this modified inequality was not achieved.

\section{All the uncertainty relations from Robertson's relation}
We show that all the uncertainty relations are derived from suitably defined Robertson's relation~\cite{fujikawa3}. We first start with Robertson's relation
\begin{eqnarray}
\sigma(M^{out}-A)\sigma(N^{out}-B)\geq \frac{1}{2}|\langle [M^{out}-A, N^{out}-B]\rangle|,
\end{eqnarray}
and the repeated use of unbiased measurement such as (1.8) gives
\begin{eqnarray}
|\langle [M^{out}-A, N^{out}-B]\rangle|
 &=&|\langle [M^{out}-A, N^{out}]\rangle|\nonumber\\
&=&|\langle [A, N^{out}]\rangle|\nonumber\\
&=&|\langle [A, N^{out}-B+B]\rangle|\nonumber\\
&=&|\langle [A, B]\rangle|\nonumber
\end{eqnarray}
where we used $[M^{out}, N^{out}]=0$. We thus obtain
\begin{eqnarray}
\sigma(M^{out}-A)\sigma(N^{out}-B)\geq \frac{1}{2}|\langle [A, B]\rangle|
\end{eqnarray}
which leads to Heisenberg error-error relation (1.11) if one recalls $\epsilon(M^{out}-A)\geq \sigma(M^{out}-A)$.

We next start with Robertson's relation  
\begin{eqnarray}
\sigma(M^{out}-A)\sigma(B^{out}-B)\geq\frac{1}{2}|\langle[M^{out}-A, B^{out}-B]\rangle|
\end{eqnarray}
and  use the triangle inequality
\begin{eqnarray}
\sigma(M^{out}-A)\sigma(B^{out}-B)&\geq& \frac{1}{2}|\langle [-A,B^{out}-B]+[M^{out}-A,-B]-[-A,-B]\rangle|\nonumber\\
&\geq& \frac{1}{2}\{|\langle [A,B]\rangle|-|\langle [A,B^{out}-B]\rangle|
-|\langle [M^{out}-A,B]\rangle|\},\nonumber\\
\end{eqnarray}
where we used $[M^{out},B^{out}]=[M,B]=0$. 
Using the variations of Robertson's relation such as
\begin{eqnarray}
\sigma(M^{out}-A)\sigma(B)\geq \frac{1}{2}|\langle [M^{out}-A,B]\rangle|
\end{eqnarray}
we obtain 
\begin{eqnarray}
\sigma(M^{out}-A)\sigma(B^{out}-B)+\sigma(M^{out}-A)\sigma(B)+
\sigma(A)\sigma(B^{out}-B)\geq \frac{1}{2}|\langle [A,B]\rangle|,
\end{eqnarray}
and
\begin{eqnarray}
\{\sigma(M^{out}-A)+\sigma(A)\}\{\sigma(B^{out}-B)+\sigma(B)\}\geq |\langle [A,B]\rangle|.
\end{eqnarray}
If one assumes the unbiased measurement and disturbance such as (1.8) in (2.4), one obtains 
\begin{eqnarray}
\sigma(M^{out}-A)\sigma(B^{out}-B)
\geq \frac{1}{2}|\langle [A,B]\rangle|.
\end{eqnarray}
We now observe that 
\begin{eqnarray}
&&\epsilon(A)\geq \sigma(M^{out}-A),\nonumber\\
&&\eta(B)\geq \sigma(B^{out}-B).
\end{eqnarray}
The quantity $\sigma(B^{out}-B)$ in Robertson's relation  is originally defined as an average of the operator $B^{out}-B$ using the state $|\psi\otimes\xi\rangle$ for any given $B^{out}$, in principle, independently of the measurement of $A$. But the quantity $\eta(B)$ is now interpreted as a disturbance caused by the measurement of $A$ by assigning suitable time dependence to $B^{out}=U^{\dagger}BU$ in the {\em Heisenberg picture}.

Thus, we obtain universally valid relations
\begin{eqnarray}
\epsilon(A)\eta(B)+\sigma(A)\eta(B)+\epsilon(A)\sigma(B)\geq \frac{1}{2}|\langle [A,B]\rangle|
\end{eqnarray}
and
\begin{eqnarray}
\left(\epsilon(A)+\sigma(A)\right)\left(\eta(B)+\sigma(B)\right)\geq |\langle [A,B]\rangle|
\end{eqnarray}
and the conditionally valid (naive) Heisenberg error-disturbance relation 
\begin{eqnarray}
\epsilon(A)\eta(B)\geq \frac{1}{2}|\langle [A,B]\rangle|.
\end{eqnarray}
Similarly, one can derive the conditionally valid Arthurs-Kelly relation
\begin{eqnarray}
\sigma(M^{out})\sigma(N^{out})\geq |\langle [A, B]\rangle|
\end{eqnarray}
using the assumption of joint unbiased measurement and the conditionally valid error-error relation in (2.2) and (1.11),
\begin{eqnarray}
\epsilon(A)\epsilon(B)\geq \frac{1}{2}|\langle [A,B]\rangle|.
\end{eqnarray}

A way to see why the naive Heisenberg error-disturbance relation
\begin{eqnarray}
\langle (M^{out}-A)^{2}\rangle^{1/2}\langle (B^{out}-B)^{2}\rangle^{1/2}\geq \frac{1}{2}|\langle [A,B]\rangle|
\end{eqnarray}
fails for spin variables is to recall that the "precise measurement" gives
\begin{eqnarray}
\langle (M^{out}-A)^{2}\rangle^{1/2}\langle (B^{out}-B)^{2}\rangle^{1/2}=0
\end{eqnarray}
irrespective of the value of $|\langle [A,B]\rangle|$.
 
The relation (2.10) corresponds to a modified form of error-disturbance relation of Ozawa~\cite{ozawa1,ozawa2}, while (2.11) gives  a  "universally valid Heisenberg uncertainty relation"~\cite{fujikawa}
\begin{eqnarray}
\bar{\epsilon}(A)\bar{\eta}(B)\geq |\langle [A,B]\rangle|
\end{eqnarray}
where
\begin{eqnarray}
"\rm inaccuracy": \ \ \ \bar{\epsilon}(A)&\equiv&\epsilon(A)+\sigma(A),  \nonumber\\
&=&\langle (M^{out}-A)^{2}\rangle^{1/2}+\langle (A-\langle A\rangle)^{2}\rangle^{1/2},\nonumber\\
"\rm fluctuation": \ \ \ \bar{\eta}(B)&\equiv&\eta(B)+\sigma(B), \nonumber\\
&=&\langle (B^{out}-B)^{2}\rangle^{1/2}+\langle (B-\langle B\rangle)^{2}\rangle^{1/2}.
\end{eqnarray} 
The definition of inaccuracy is natural for the precise measurement with $\epsilon(A)=0$ of a broadly spread state: $\langle (M^{out}-A)^{2}\rangle^{1/2}=0$ implies $M^{out}|\psi\rangle\otimes|\xi\rangle=A|\psi\rangle\otimes|\xi\rangle$ and thus the residual fluctuation in the apparatus is given by $\sigma(M^{out})=\sigma(A)$.

The relation (2.11) is interesting in comparison with an  error-disturbance version of Arthurs-Kelly relation. The (modified) Arthurs-Kelly relation in this context is defined using (2.12) as
\begin{eqnarray}
&&\sigma^{2}(M^{out})\sigma^{2}(B^{out})\nonumber\\
&&=\{\langle (M^{out}-A)^{2}\rangle+\sigma^{2}(A)\}\{\langle (B^{out}-B)^{2}\rangle+\sigma^{2}(B)\}\nonumber\\
&&\geq\frac{1}{4}|\langle [A,B]\rangle|^{2}\{ \langle (B^{out}-B)^{2}\rangle^{-1}+\sigma^{-2}(B)\}\{\langle (B^{out}-B)^{2}\rangle+\sigma^{2}(B)\}\nonumber\\
&&\geq|\langle [A,B]\rangle|^{2}.
\end{eqnarray}
If one recalls the relation
\begin{eqnarray}
\bar{\epsilon}(A)\bar{\eta}(B)&=&\{\epsilon(A)+\sigma(A)\}
\{\eta(B)+\sigma(B)\}\nonumber\\
&\geq&\{\langle (M^{out}-A)^{2}\rangle+\sigma^{2}(A)\}^{1/2}\nonumber\\
&&\times\{\langle (B^{out}-B)^{2}\rangle+\sigma^{2}(B)\}^{1/2}\nonumber\\
&=&\sigma(M^{out})\sigma(B^{out}), 
\end{eqnarray}
the modified Arthurs-Kelly relation (2.19) is more restrictive than the universally valid Heisenberg relation (2.11).
The comparison of these two relations with spin measurement~\cite{hasegawa} illustrates that uncertainty relations are very subtle. The stronger modified Arthurs-Kelly relation totally fails while the universally valid Heisenberg relation
cannot achieve the lower bound. This is shown in Fig.1 which is taken from Ref.~\cite{fujikawa3}. 
\begin{figure}[h]
  \begin{center}
    \includegraphics[width=140mm]{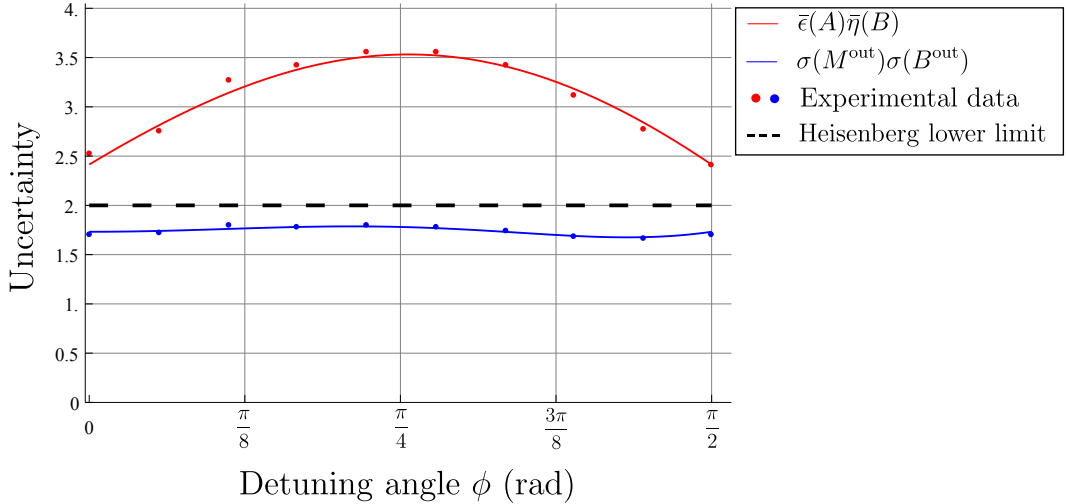}
  \end{center}
  \caption{{\bf Experimental test of the universally valid Heisenberg uncertainty relation (red line) and the modified Arthurs-Kelly relation  (blue line):}
  }
  \label{.eps}
\end{figure}

We may thus conclude that all the universally valid relations (in term of $\eta$ and $\epsilon$) are the secondary consequences of Robertson's relation.
Also, the saturation of Robertson's relation is a {\em necessary condition} of the saturation of universally valid uncertainty relations. In fact,
Robertson's relation (2.3) is always saturated for the precise measurement while modified versions such as (2.10) and (2.11) are not saturated in general.

\section{Algebraic inconsistency of unbiasedness conditions}

The unbiased measurement implies the relations
\begin{eqnarray}
\langle M^{out}- A\rangle=0, \ \ \ \ \langle N^{out}- B\rangle=0
\end{eqnarray}
for {\em all} the states $\psi$ in $|\psi\otimes \xi\rangle$; $|\psi\rangle$ and $|\xi\rangle$ stand for the states of the system and the apparatus. 
$M^{out}=U^{\dagger}(1\otimes M)U$ and $N^{out}=U^{\dagger}(1\otimes N)U$ stand for the variables after the measurement. The unitary operator $U=U(t_{\rm final})=\exp\{-\frac{i}{\hbar}\hat{H}t_{\rm final}\}$ generates  the measurement interaction, and we assume $U(t)=U(t_{\rm final})$ for $t>t_{\rm final}$.  
The precise measurement implies,
\begin{eqnarray}
\langle (M^{out}- A)^{2}\rangle=0 \ \  {\rm or} \ \  (M^{out}- A)|\psi\otimes \xi\rangle=0
\end{eqnarray}
for arbitrary given operator $A$ and state $|\psi\rangle$.   
 
We assume $[M, N]=0$, and thus for the unitary time development 
$\langle [M^{out}, N^{out}]\rangle=0$.
This relation when combined with the precise measurement of $A$ implies
\begin{eqnarray}
\langle [M^{out}, N^{out}]\rangle&=&\langle [A, N^{out}]\rangle\nonumber\\
&=&\langle [A, N^{out}-B]\rangle+\langle [A, B]\rangle\nonumber\\
&=&\langle [A, B]\rangle
\end{eqnarray}
where we used the relation (1.8), 
\begin{eqnarray}
\langle [A, N^{out}-B]\rangle=0.
\end{eqnarray}
We thus conclude~\cite{fujikawa2}
\begin{eqnarray}
\langle [A, B]\rangle=\langle [M^{out}, N^{out}]\rangle=0
\end{eqnarray}
which is a contradiction since $\langle [A, B]\rangle\neq 0$ in general.   

This relation  may be interpreted that the precise measurement of $A$ does not allow the unbiased measurement of $B$, if all the operators involved are assumed to be {\em well-defined}.

Similarly, one concludes~\cite{fujikawa2} 
\begin{eqnarray}
\langle [A, B]\rangle=\langle [M^{out}, B^{out}]\rangle=0
\end{eqnarray}
if one assumes the precise measurement of $A$ and the unbiased disturbance of 
$B$ which implies $\langle B^{out} - B \rangle=0$ for all $\psi$. Here $B^{out}=U^{\dagger}(B\otimes 1)U$ 
stands for the variable $B$ after the {\em measurement} of $A$. Note that   $[M^{out}, B^{out}]=[M, B]=0$.

We interpret the algebraic inconsistency (3.6) as an indication of the failure of the assumption of unbiased disturbance of $B$ for the precise projective measurement of $A$, if all the operators involved are {\em well-defined}. \\

We now briefly discuss the implications of these mathematical analyses on uncertainty relations:\\
1. The Heisenberg error-error relation 
\begin{eqnarray}
\epsilon(M^{out}-A)\epsilon(N^{out}-B)\geq \frac{1}{2}|\langle [A,B]\rangle|
\end{eqnarray}
and the Arthurs-Kelly relation
\begin{eqnarray}
\sigma(M^{out})\sigma(N^{out})\geq |\langle [A, B]\rangle|
\end{eqnarray}
are expected to be valid as conditionally valid uncertainty relations. In this case the apparatus variable $N^{out}$ becomes {\em singular} for the precise measurement of $A$, namely, for $M^{out}-A\rightarrow 0$ if the unbiasedness condition $\langle N^{out}-B \rangle=0$ is imposed: This conclusion is based on the recent numerical analysis in the framework of quantum estimation theory~\cite{watanabe1, watanabe2}. 
\\
2. Naive Heisenberg error-disturbance relation
\begin{eqnarray}
\epsilon(M^{out}-A)\eta(B^{out}-B)\geq \frac{1}{2}|\langle [A,B]\rangle|
\end{eqnarray}
fails for the bounded operators;
$\eta(B^{out}-B)$ is bounded (and thus well-defined), and consequently this relation fails for $\epsilon(M^{out}-A)\rightarrow 0$.

\section{ Discussion and conclusion}

\noindent
(i)We have shown that all the uncertainty relations in terms of $\epsilon$ and $\eta$ are derived from Robertson's relation. 
\\
\\
(ii)We have also shown that the joint unbiased measurements or unbiased measurement and unbiased disturbance are algebraically inconsistent in general independently of uncertainty relations.  
\\

I presented my view of the uncertainty relation, in particular,
I emphasized in my talk that the uncertainty relation should be simple and understandable to every undergraduate if it should be fundamental. Of course, the subject of the uncertainty relation is very old and there exist many different points of view. I just mention some of the past works~\cite{past1,past2,past3,past4,past5} which I found interesting, in addition to the most recent activities on this subject~\cite{new1,new2,new3,new4,new5,new6,new7}. 
\\

Finally, we mention an amusing  relation for the {\em box normalization} in quantum mechanics. In the interval $-\frac{L}{2}\leq x \leq \frac{L}{2}$ with $\psi(-\frac{L}{2},t)=\psi(\frac{L}{2},t)$, we have~\cite{fujikawa3}
\begin{eqnarray}
\Delta p\Delta x\geq\frac{\hbar}{2}|1-L|\psi(\frac{L}{2},t)|^{2}|
\end{eqnarray}
which does not appear to be widely recognized. 

This relation is regarded as a $px$ version of the universally valid Kennard-Robertson uncertainty relation.
Note that the momentum eigenstate with the periodic boundary condition in a finite box gives $\Delta(p)=\langle (p-\langle p\rangle)^{2}\rangle^{1/2}=0$ with a finite value for $\Delta(x)=\langle (x-\langle x\rangle)^{2}\rangle^{1/2}$, and thus the left-hand side of this relation vanishes. The right-hand side of (4.1) also vanishes in this case and thus it is consistent.

\end{document}